\documentclass[aps,prl,twocolumn,showpacs,amsmath,amssymb, nofootinbib, noeprint]{revtex4-1}
\usepackage{epsfig}
\usepackage{booktabs}
\usepackage{subfigure}
\usepackage{graphicx}% Include figure files
\usepackage{dcolumn}% Align table columns on decimal point
\usepackage{bm}% bold math
\usepackage{xcolor}
\usepackage{float}
\definecolor{aa}{RGB}{0,0,139}

\usepackage[colorlinks,urlcolor=aa,linkcolor=blue,anchorcolor=aa,citecolor=aa]{hyperref}
\usepackage[mathlines]{lineno}% Enable numbering of text and display math
\usepackage{multirow}
\usepackage{enumerate}
\usepackage{amsmath}
\usepackage{tabularx}
\usepackage{relsize}
\def\babar{\mbox{\slshape B\kern-0.1em{\smaller A}\kern-0.1em
    B\kern-0.1em{\smaller A\kern-0.2em R}}}

\newcommand{\BESIIIorcid}[1]{\href{https://orcid.org/#1}{\hspace*{0.1em}\raisebox{-0.45ex}{\includegraphics[width=1em]{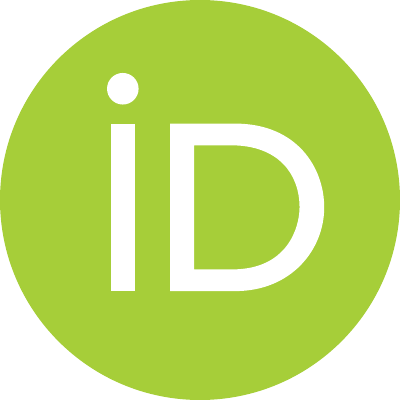}}}}

\begin{document}
% \linenumbers
\preprint{}

\title{{\bf \boldmath First Observation of the Radiative Decay $D^0 \to \gamma\omega$}}

\author{
\begin{small}
  \begin{center}
%% Saved at => 2026-03-11
M.~Ablikim$^{1}$\BESIIIorcid{0000-0002-3935-619X},
M.~N.~Achasov$^{4,c}$\BESIIIorcid{0000-0002-9400-8622},
P.~Adlarson$^{83}$\BESIIIorcid{0000-0001-6280-3851},
X.~C.~Ai$^{89}$\BESIIIorcid{0000-0003-3856-2415},
C.~S.~Akondi$^{31A,31B}$\BESIIIorcid{0000-0001-6303-5217},
R.~Aliberti$^{39}$\BESIIIorcid{0000-0003-3500-4012},
A.~Amoroso$^{82A,82C}$\BESIIIorcid{0000-0002-3095-8610},
Q.~An$^{78,65,\dagger}$,
Y.~H.~An$^{89}$\BESIIIorcid{0009-0008-3419-0849},
M.~S.~Anderson$^{39}$\BESIIIorcid{0009-0008-1550-2632}
% ...
\\
\vspace{0.2cm}
(BESIII Collaboration)\\
\vspace{0.2cm} {\it
$^{1}$ Institute of High Energy Physics, Beijing 100049, People's Republic of China\\
% ...
$^{4}$ Budker Institute of Nuclear Physics SB RAS (BINP), Novosibirsk 630090, Russia\\
% ...
$^{31}$ INFN Sezione di Ferrara, (A)INFN Sezione di Ferrara, I-44122, Ferrara, Italy; (B)University of Ferrara, I-44122, Ferrara, Italy\\
% ...
$^{39}$ Johannes Gutenberg University of Mainz, Johann-Joachim-Becher-Weg 45, D-55099 Mainz, Germany\\
% ...
$^{65}$ State Key Laboratory of Particle Detection and Electronics, Beijing 100049, Hefei 230026, People's Republic of China\\
% ...
$^{78}$ University of Science and Technology of China, Hefei 230026, People's Republic of China\\
% ...
$^{82}$ University of Turin and INFN, (A)University of Turin, I-10125, Turin, Italy; (B)University of Eastern Piedmont, I-15121, Alessandria, Italy; (C)INFN, I-10125, Turin, Italy\\
$^{83}$ Uppsala University, Box 516, SE-75120 Uppsala, Sweden\\
% ...
$^{89}$ Zhengzhou University, Zhengzhou 450001, People's Republic of China\\

\vspace{0.2cm}
$^{\dagger}$ Deceased\\
% ...
$^{c}$ Also at the Novosibirsk State University, Novosibirsk, 630090, Russia\\
% ...

}
%% ends here %%

\end{center}
\end{small}
}

% \vspace{0.4cm}
% \date{\today}
%------------------------------------------------------------------------------
\begin{abstract}

We report the first observation of the radiative charm decay $D^0 \to \gamma \omega$ with a significance of $6.0\sigma$.  The analysis uses 20.3\,$\rm{fb}^{-1}$ of $e^+e^-$ collision data collected by the BESIII detector at a center-of-mass energy of 3.773\,GeV, and employs a deep learning strategy based on the Particle Transformer architecture.  The absolute branching fraction of $\mathcal{B}(D^0\to\gamma\omega)=(2.12\pm 0.40_{\rm stat}\pm 0.25_{\rm syst})\times 10^{-5}$ confirms the dominance of long-distance hadronic dynamics in radiative charm decays and provides a valuable benchmark for testing non-perturbative QCD calculations of $D \to \gamma V$ transitions.
\end{abstract}

%\pacs{Valid PACS appear here}
\maketitle

\begin{figure}[htbp]
  \vspace{-0.5em}
  \centering
          \mbox{
            \put(-128, 0){\includegraphics[width=0.25\textwidth]{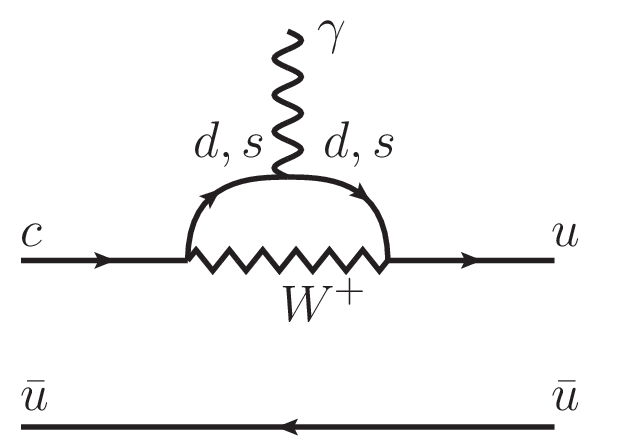}
            }
            \put(0, 0){\includegraphics[width=0.25\textwidth]{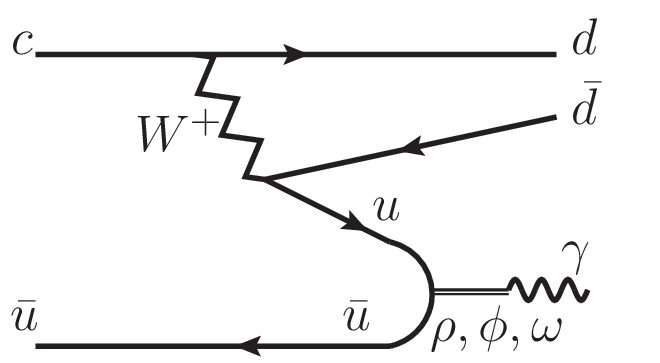}
            }
            \put(-80, -8) { $(a)$}
            \put(60, -8)  {$(b)$}
        }
  \caption{The decay mechanism diagrams illustrating representative contributions to the radiative charm decay: (a) a short-distance amplitude, and (b) a long-distance VMD amplitude.  Each diagram shows one of several possible topologies within its respective mechanism. Other diagrams contributing to related decays are discussed in Refs.~\cite{Burdman:1995te,Cao:2023gfv}. 
  }
  \label{fig:feynman_diagram}
\end{figure}

%\tableofcontents
%------------------------------------------------------------------------------
The flavor-changing neutral current process of radiative charm decay, $c\to \gamma u$, is highly suppressed in the Standard Model due to the Glashow-Iliopoulos-Maiani mechanism and the smallness of the relevant Cabibbo-Kobayashi-Maskawa matrix elements. 
The $c\to \gamma u$ transition proceeds via the electromagnetic penguin diagram, as depicted in Fig.~\ref{fig:feynman_diagram}(a), representing a short-distance mechanism in which the photon is emitted via the weak interaction.
Theoretical calculations for its contribution to $D\to \gamma V$ decays, where $V$ denotes a vector meson with matching quark content, predict branching fractions~(BFs) of~$\mathcal{O}(10^{-12}\text{--}10^{-9})$~\cite{Burdman:1995te, Greub:1996wn, Bajc:1994ui, Cheng:1994kp}, which are several orders of magnitude lower than the experimental measurements~\cite{PDG}.
%Theoretical calculations for its contribution to the
%branching fraction~(BF) of the Cabibbo-suppressed $D\to \gamma V$ decays, where
%$V$ denotes a vector meson with the appropriate quark composition, has beed
%studied with various approaches.
%including modified quark model~\cite{Asthana:1990kr}, effective Lagrangian~\cite{Cheng:1994kp},
%the heavy quark effective field theory and chiral Lagrangian~\cite{Bajc:1994ui}.
%These calculations predict BFs in the
%range~$\mathcal{0}(10^{-8}\text{--}10^{-9})$, which are systematically lower
%than the experimental measurements by several orders of magnitude~\cite{Burdman:1995te, Greub:1996wn, Bajc:1994ui, Cheng:1994kp, Bajc:1994ui}.

% \begin{figure}[htbp]
%   \vspace{-2.5em}
%   \centering
  
% \hspace*{-2em}
% \includegraphics[width=0.35\textwidth]{Feynman_diagram_short_effect.pdf}
  
% \hspace*{-2em}
% \includegraphics[width=0.4\textwidth]{Feynman_diagram_long_effect.pdf}

% % 调整标签位置
% \hspace*{-8em}\raisebox{-2em}{(a)} % 左图标签
% \hspace*{4em}\raisebox{-2em}{(b)}  % 右图标签
        
%   \caption{Feynman diagrams for the decay $D^0 \to \gamma \omega$, which is mediated via (a) short-distance processes and (b) long-distance VMD processes.
%   }
%   \label{fig:feynman_diagram}
% \end{figure}

%------------------------------------------------------------------------------
This discrepancy in the BF could be resolved through electromagnetic effects induced via the vector-meson dominance~(VMD) mechanism, a long-distance process where the photon is generated away from the weak interaction vertex. 
Various quantum chromodynamics-based~(QCD-based) calculations have shown that the VMD mechanism could enhance the BFs to
$\mathcal{O}(10^{-6}\text{--}10^{-4})$~\cite{Cao:2023gfv, Fajfer:1997bh, Fajfer:1998dv, deBoer:2017que, Biswas:2017eyn, Khodjamirian:1995uc, Shen:2013oua}.
In the $b$ sector, it is known that the VMD mechanism yields a non-dominant but non-negligible contribution~\cite{Cheng:1994qw, Ali:1995uy, Donoghue:1996fv}. 
Testing VMD in radiative charm decays provides crucial constraints on its role in the
radiative $B$ decay sector.
In the VMD framework, $D\to \gamma V$ decays receive contributions from $D\to VV^{\prime}$ processes where $V^{\prime}$ converts to a photon; one possible diagram is shown in Fig.~\ref{fig:feynman_diagram}(b). 
Consequently, the BFs of $D\to \gamma V$ and $D\to VV^{\prime}$ are linked through the decay constant of the vector-meson $V^{\prime}$ and its transverse polarization fraction, reflecting the purely transverse polarization of a real photon. 

The $D\to \gamma V$ decays can serve as sensitive probes for new physics~\cite{deBoer:2017que,Biswas:2017eyn,deBoer:2018zhz,LHCb:2014vnw}. The photons from the short-distance electromagnetic penguin contribution decay will be mostly left-handed. Even in the presence of a long-distance contribution, an unusually large fraction of right-handed photon polarization in the $D^0\to \gamma\omega$ decay would provide a clear signature of new physics~\cite{deBoer:2018zhz}. 
Another approach, as illustrated in Ref.~\cite{Fajfer:2000zx}, is to utilize the unequal contribution from long-distance amplitudes to the $\gamma(u\bar{u})$ and $\gamma(d\bar{d})$ final states.
Since the $\omega$ meson is an isospin singlet, $\frac{1}{\sqrt{2}}(u\bar{u}+d\bar{d})$, and the $\rho^0$ meson is an isospin triplet, $\frac{1}{\sqrt{2}}(u\bar{u}-d\bar{d})$, a comparison of the decay rates for $D^0\to \gamma\omega$ and $D^0\to \gamma\rho^0$ can effectively isolate the short-distance contribution, thereby enabling searches for new physics effects.
To this end, Ref.~\cite{Fajfer:2000zx} introduces the observable $\mathcal{D}^{\omega-\rho}$, defined as \begin{equation}
\resizebox{.9\hsize}{!}{$
\mathcal{D}^{\omega-\rho} \equiv \frac{
  \Gamma(D^0 \to \gamma\omega)/(m_D^2 - m_\omega^2)^3 
  \, - \,  
  \Gamma(D^0 \to \gamma\rho^0)/(m_D^2 - m_{\rho^0}^2)^3
}{
  \Gamma(D^0 \to \gamma\omega)/(m_D^2 - m_\omega^2)^3
}$},\end{equation}
\noindent where $\Gamma$ and $m$ denote the partial decay width and mass of the corresponding meson, respectively.
Given its unique importance, the decay $D^0 \to \gamma\omega$ has been extensively studied theoretically. 
Short-distance calculations with two-loop QCD corrections predict a BF of $\mathcal{O}(10^{-9})$~\cite{deBoer:2017que}, 
whereas long-distance models—including hard spectator interaction (HSI), weak annihilation (WA), hybrid frameworks, the covariant light-front quark model (CLFQM), and VMD—yield predictions of $\mathcal{O}(10^{-7}\text{--}10^{-5})$~\cite{deBoer:2017que,Fajfer:1998dv,Burdman:1995te,Cao:2023gfv,CLEO:1998mtp}.

However, radiative charm decays are experimentally challenging due to overwhelming background contamination from hadronic decays of charm mesons, primarily because photons from $\pi^0$ or $\eta$ decays are often misidentified as signal radiative photons. 
Consequently, experimental measurements remain scarce: while decays such as $D^0\to\gamma\phi$, $D^0\to\gamma\bar{K}^{*0}$, and $D^0\to\gamma\rho^0$ have been observed by the Belle~\cite{Belle:2003vsx,Belle:2016mtj} and BaBar~\cite{BaBar:2008kjd} collaborations,
only an upper limit of $\mathcal{B}(D^0\to\gamma\omega)< 2.4 \times 10^{-4}$ at the 90\% confidence level has been reported by the CLEO collaboration~\cite{CLEO:1998mtp}.

As established in Refs.~\cite{BESIII:2017whk, Belle:2016mtj}, traditional cut-based selection methods fail to construct effective discriminators to suppress these backgrounds. 
This limitation motivates the adoption of a deep neural network (DNN)~\cite{LeCun:2015pmr}, which is trained to classify signal and background events and provides unprecedented discrimination power. 
The DNN processes reconstruction-level information from tracks and showers, automatically exploiting their underlying correlations and extracting discriminative patterns beyond those accessible through conventional analysis methods~\cite{BESIII:2024zfv,BESIII:2024thr}.

In this Letter, we present the first observation of the radiative decay $D^0\to\gamma\omega$ using $e^+e^-$ collision data collected with the BESIII detector at $\sqrt{s}=3.773$\,GeV corresponding to an integrated luminosity of 20.3\,\ensuremath{\mathrm{fb}^{-1}}~\cite{Ablikim:2013ntc, BESIII:2015equ, BESIII:2024lbn}.
The analysis employs a DNN model for binary signal–background classification, enabling the significant background suppression required for this observation.
Charge-conjugate modes are implied throughout.

A description of the design and performance of the BESIII detector can be found in Ref.~\cite{BESIII:2009fln,Guo:2017sjt,Li:2017jpg,Cao:2020ibk,BESIII:2020nme}. Simulated data samples are produced with a {\sc geant4}-based~\cite{GEANT4:2002zbu} Monte Carlo~(MC) toolkit, which includes the geometric description~\cite{Huang:2022wuo} of the BESIII detector and the detector response. The simulation models the beam energy spread and initial state radiation (ISR) in the $e^+e^-$ annihilations with the generator {\sc kkmc}~\cite{Jadach:2000ir}. The inclusive MC sample includes the production of $D\bar{D}$ pairs, the non-$D\bar{D}$ decays of the $\psi(3770)$, the ISR production of the $J/\psi$ and $\psi(3686)$ states, and the continuum processes incorporated in {\sc kkmc}. All particle decays are modeled with {\sc evtgen}~\cite{Lange:2001uf, Ping:2008zz} using BFs either taken from the Particle Data Group~(PDG)~\cite{PDG}, when available, or otherwise estimated with {\sc lundcharm}~\cite{Chen:2000tv, Yang:2014vra}. Final state radiation from charged final state particles is incorporated using {\sc photos}~\cite{Richter-Was:1992hxq}.
The signal MC samples
for the decay $D^0\to\gamma\omega$ are simulated with a generator developed using helicity amplitudes, a model previously employed in the
measurements conducted by Belle~\cite{Belle:2016mtj} and BaBar~\cite{BaBar:2008kjd}.

%\section{Method}
%------------------------------------------------------------------------------
The double-tag~(DT) method~\cite{MARK-III:1985hbd} provides high-purity samples for measuring the absolute BFs of $D$ meson decays, eliminating the dependence on the integrated luminosity and the $D\bar{D}$ production cross section~\cite{BESIII:2024zvp}. 
The tagged $\bar{D}^0$ mesons are first reconstructed in one of the three decay modes $ K^+ \pi^-$, $ K^+ \pi^- \pi^0$, and $ K^+ \pi^- \pi^+ \pi^-$, referred to as single-tag~(ST) events. 
An event in which both a signal decay and an ST $\bar{D}^0$ are simultaneously reconstructed is referred to as a DT event.
The absolute BF of the signal decay is determined by
  ${\mathcal B}_{{\rm sig}} = N_{\rm DT}/(N^{\rm tot}_{\rm ST} \, \epsilon_{{\rm sig}} \, \mathcal{B}_{\rm sub})$,
% \begin{equation}
%   \label{eq:br}
%   {\mathcal B}_{{\rm sig}} = N_{\rm DT}/(N^{\rm tot}_{\rm ST}\cdot\epsilon_{{\rm sig}}),
% \end{equation}
where $N^{\rm tot}_{\rm ST}=\sum_\alpha N_{{\rm ST}}^\alpha$ is the total ST yield, and $N_{\rm DT}$ is the DT yield in data. 
Here, $N_{{\rm ST}}^\alpha$ denotes the  ST yield for the tag mode $\alpha$, and the signal detection efficiency $\epsilon_{{\rm sig}}$ is averaged over the tag modes,
  $\epsilon_{\rm sig} = \frac{\sum_\alpha (N^\alpha_{\rm ST} \, \epsilon^\alpha_{\rm DT}/\epsilon^\alpha_{\rm ST})}{N^{\rm tot}_{\rm ST}}.$
% \begin{equation}  
% \label{br}
%   \epsilon_{\rm sig} = \frac{\sum_i (N^i_{\rm ST}\cdot \epsilon^i_{\rm DT}/\epsilon^i_{\rm ST})}{N^{\rm tot}_{\rm ST}}.
% \end{equation}
The quantities $\epsilon^\alpha_{\rm ST}$ and $\epsilon^\alpha_{ \rm DT}$ are the ST and DT efficiencies for the tag mode $\alpha$.
Finally, $\mathcal{B}_{\rm sub}$ represents the product of the BFs of all intermediate particles in the signal decay, $\mathcal{B}_{\rm sub}=\mathcal{B}_{\omega\to\pi^{+}\pi^{-}\pi^{0}}\mathcal{B}_{\pi^{0}\to\gamma\gamma}$~\cite{PDG}.

%------------------------------------------------------------------------------
The selection criteria for 
% the final state particles
$K^{\pm}$, $\pi^{\pm}$, $\gamma$ and $\pi^0$ candidates
are the same as {
% \color{red}
Ref.~\cite{BESIII:2024ncc}.}
The tagged $\bar D^0$ mesons are reconstructed using the energy difference $\Delta E \equiv  E_{\bar D^0}-E_{\rm beam}$ and the beam-constrained mass $M_{\rm BC} \equiv \sqrt{E^{2}_{\rm beam}/c^{4}-|\vec{p}_{\bar D^0}|^{2}/c^{2}}$, where $E_{\rm beam}$ is the beam energy, and $\vec{p}_{\bar D^0}$ and $E_{\bar D^0}$ are the  measured momentum and energy of the $\bar D^0$ candidate in the $e^+e^-$ rest frame, respectively.
If multiple candidates are found in an event, the one with the minimum $|\Delta{E}|$ is chosen. 
The ST yields and efficiencies are determined by fitting the $M_{\rm BC}$ distributions of the accepted candidates in data and inclusive MC sample, respectively. The ST yields in the data and the corresponding ST efficiencies are listed in Table~\ref{tab:signal efficiency}. 
More details about the ST selection can be found in {
% \color{red}
Ref.~\cite{BESIII:2025hdt}
}.
Summing over all three tag modes, the total ST yield within the $M_{\rm BC}$ signal region, from 1.859 to 1.873~$\textrm{GeV}/c^{2}$, is $N_{\rm ST}^{\rm tot} = (16135.4\pm4.6)\times10^3$.

\begin{table}[htbp]
\vspace{-1em}
    \caption{The energy difference~($\Delta E$) windows, ST yields in data~($N_\textrm{ST}^{\alpha}$), ST efficiencies~($\epsilon_\textrm{ST}^{\alpha}$), and DT efficiencies~($\epsilon_\textrm{DT}^{\alpha}$) for each tag mode $\alpha$. These efficiencies do not include the BFs of all possible intermediate particles.The uncertainties shown are statistical only.}
    \label{tab:signal efficiency}
    \resizebox{1.0\linewidth}{!}{
    \begin{tabular} {l | c c c c}
        \hline \hline
        Tag mode & $\Delta E$~(MeV) & $N_{\rm{ST}}^{\alpha}~(\times 10^{3})$  & $\epsilon_{\rm{ST}}^{\alpha}(\%)$ & $\epsilon_{\rm{DT}}^{\alpha}(\%)$ \\
        \hline
$K^+ \pi^-$ & $[-27, 27]$ & $3725.7 \pm 2.0$ & $65.10 \pm 0.01$ & $18.79 \pm 0.05$ \\$K^+ \pi^- \pi^0$ & $[-62, 49]$ & $7422.3 \pm 3.2$  &  $35.60 \pm 0.01$ & $9.38 \pm 0.02$ \\$K^+ \pi^- \pi^+ \pi^-$ & $[-26, 24]$ & $4987.5 \pm 2.5$ &  $40.94 \pm 0.01$ & $10.19 \pm 0.02$ \\
        \hline \hline
    \end{tabular}
    }
\end{table}

% Once a ST $\bar{D}^0$ decay is identified, a search is performed for a $D^0$ signal decay in the same event.
After an  ST $\bar{D}^0$ is identified, the signal decay $D^{0} \to \gamma \omega$ is reconstructed on the recoil side.
The most energetic photon candidate passing the selection criteria described in Ref.~\cite{BESIII:2024ncc} is selected as the radiative photon.
Then, we require exactly two oppositely charged pions.
To maintain high efficiency, multiple $\pi^0$ candidates are allowed.  
For those events with multiple candidates, the $\pi^0$ meson forming an invariant mass $M_{\pi^+\pi^-\pi^0}$ closest to the known $\omega$ mass~\cite{PDG} is selected. 
Prior to training the DNN, we apply loose kinematic constraints to remove background events that lie far from the signal region in the high-dimensional space, while ensuring negligible loss of signal efficiency.
The invariant mass $M_{\pi^+\pi^-\pi^0}$ is required to lie in the range $(0.65, 0.90)~\mathrm{GeV}/c^2$, 
% which corresponds to a loose mass window that fully covers the $\omega$ peak structure.
which corresponds to about $\pm 10\sigma$ around the known $\omega$ mass~\cite{PDG}, where $\sigma$ is the mass resolution of $M_{\pi^+\pi^-\pi^0}$ from the signal MC sample.
To suppress the background from $D^0 \to \pi^+\pi^-\pi^0 K_L^0$ with a misidentified photon, the recoil mass-squared against the $\pi^+\pi^-\pi^0$ system is required to be less than $0.2~\mathrm{GeV}^2/c^4$.
% ($|RM^{2}_{\pi/^+\pi^-\pi^0}|<0.2$~GeV/$c^2$).
On the signal side, $M_{\rm BC}^{\rm sig}$ and $\Delta E_{\rm sig}$ are required to satisfy $1.83 < M_{\rm BC}^{\rm sig} < 1.89~\mathrm{GeV}/c^2$ and $-0.1 < \Delta E_{\rm sig} < 0.1~\mathrm{GeV}$, respectively.

%------------------------------------------------------------------------------
In processing the $e^+e^-$ collision data with the DNN, we conceptualize the reconstructed particles in an event as a ``point cloud''~\cite{Qu:2019gqs}, i.e., an unordered and variable-sized set of points in a high-dimensional feature space. 
This representation is fed into a Transformer-based model architecture, Particle Transformer~\cite{pmlr-v162-qu22b}, renowned for its state-of-the-art performance in jet tagging~\cite{Larkoski:2017jix}.
% ParT has been widely adopted in BESIII analyses~\cite{BESIII:2024mgg,BESIII:2024cbr,BESIII:2025biy}.
In this Letter, the dataset for training DNN is prepared as a random shuffle of signal and background events, with equal statistics after event selection.
The signal events are from the signal MC sample, while the background events are taken from the inclusive MC sample, which contains all known $D\bar{D}$ decays but no signal process. 
The dataset comprises approximately 0.84 million events in total, with 80\% used for training and the remaining 20\% reserved for independent validation.

The input information from an event to the DNN includes features of all charged tracks reconstructed in the main drift chamber (MDC) and isolated showers clustered in the electromagnetic calorimeter, forming two sets of ``particle clouds''. 
For every charged track, the features comprise the azimuthal and polar angles in the laboratory frame, magnitude of momentum, and parameters characterizing the helical trajectory in the MDC. Additionally, normalized differences between the measured and predicted specific ionization energy loss d$E$/d$x$ for pion hypotheses, along with similar differences for time-of-flight measurements, are used to summarize particle identification information. 
For each shower, the features include the azimuthal and polar angles, energy deposition, number of crystals above threshold, energy ratio between $3\times3$ and $5\times5$ crystal groupings surrounding the shower center, lateral and secondary moments, and the A20 and A42 Zernike moments~\cite{Sinkus:1996ch}. 
Data and MC simulations are consistent for these adopted features.
Subsequently, the full set of hyperparameters (including network architecture~\cite{Qu:2019gqs}, regularization strength, and optimization settings~\cite{BESIII:2024mgg,BESIII:2024cbr,BESIII:2025biy}) is optimized to achieve optimal performance while avoiding overfitting.

% The model hyperparameters have been optimized to maximize performance while preventing overfitting. 
% The adapted architecture comprises four particle attention blocks and two class attention blocks. 
% Particle features are first embedded using a three-layer multilayer perceptron with hidden dimensions of 64, 128, 64, while pairwise interaction features are encoded via a four-layer point-wise 1D convolutional network with channel dimensions 64, 64, 64, and 8. A dropout rate of 0.1 is applied throughout the network to mitigate overfitting. 
% The DNN was trained on the aforementioned dataset for 50 epochs with a batch size of 512.

%------------------------------------------------------------------------------
Since the dominant hadronic background $D^0 \to \pi^+ \pi^- \pi^0 \pi^0$ differs from the signal by one extra photon, it exhibits shapes nearly identical to the signal in both the $M_{\rm BC}^{\rm sig}$ and $\Delta E_{\rm sig}$ distributions, rendering these variables ineffective for signal yield extraction. 
In contrast, the DNN outputs a classification score in the range $[0,1]$, interpreted as the probability that an event is signal-like, which provides significantly improved discrimination between signal and background. 
We therefore perform an unbinned maximum likelihood fit to the DNN score distribution of data to extract the DT yield $N_{\rm DT}$. 
The fit is restricted to the interval $[0.5,\,1.0]$, where the signal events are concentrated, to ensure the signal component stands out clearly.
The signal and background shapes in the fit are extracted from the signal MC sample and the inclusive MC sample, respectively.
The corresponding DT efficiencies are presented in Table~\ref{tab:signal efficiency} and then the BF can be determined.

\begin{figure}[htpb]
\centering
% \vspace{-2em}
\includegraphics[width=0.47\textwidth]{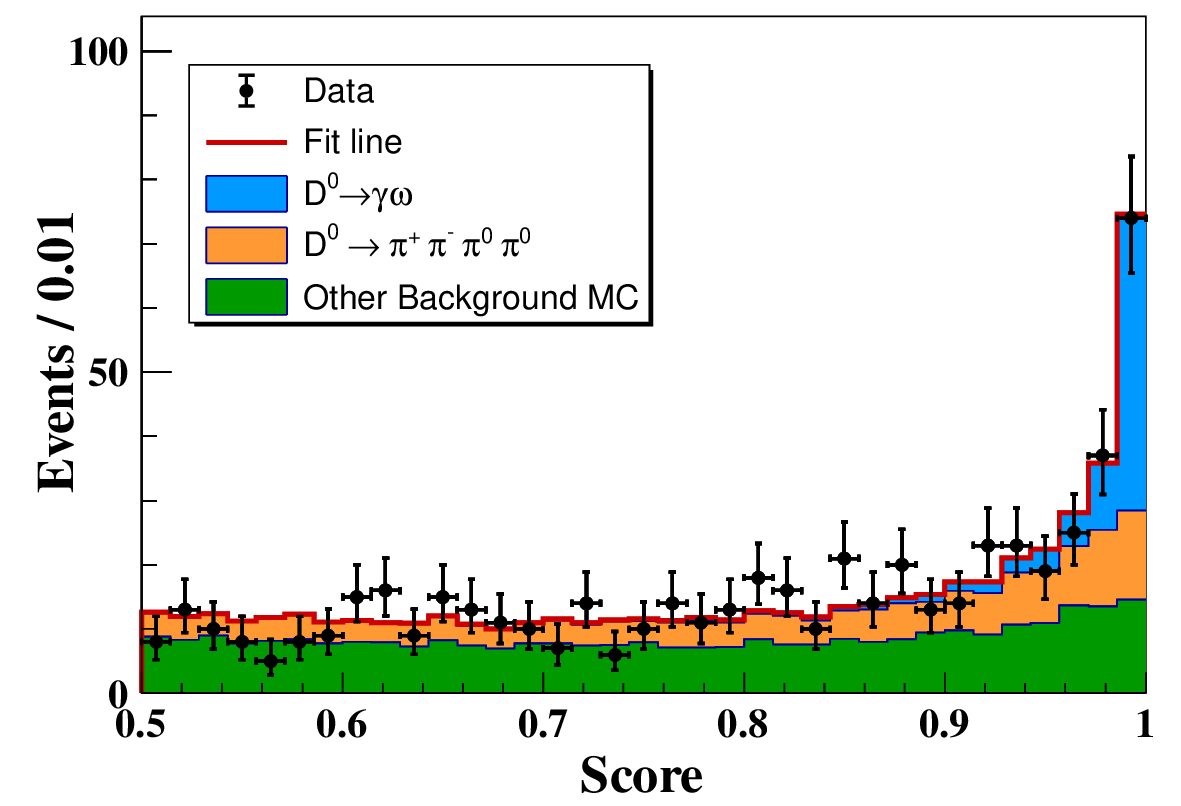}
\caption{Fit to the DNN output score distribution. The points with error bars are the data. The red curve represents the total fit, while the blue, orange, and green histograms show the signal and the two background components.}
\label{fig:fit score}
\end{figure}

In addition, a machine-learning technique known as ``model ensemble''~\cite{Mehta:2018dln} is employed to enhance robustness: 100 individual DNNs are trained in parallel with random variations in network initialization, batch processing sequence, and dropout mechanism~\cite{srivastava2014dropout}. 
% By averaging their outputs during inference, we construct an ensemble DNN that exhibits greater robustness and generalization ability than any single trained model. 
Repeating the aforementioned prediction and fitting procedure 100 times yields a distribution of BFs, which is fitted with a Gaussian function. 
The model associated with the BF nearest to the Gaussian mean is adopted as the nominal model~ (see the supplemental material~\cite{supplement} for details).
The nominal fit is shown in Fig.~\ref{fig:fit score}, where a clear enhancement is observed in the high-score region.
We obtain $N_{\rm DT}=79.8\pm14.9$ with a statistical significance of $7.0\sigma$, without including systematic effects.
The statistical significance is evaluated as $\sqrt{-2 \ln(\mathcal{L}_{0}/\mathcal{L})}$,
where $\mathcal{L}$ denotes the maximum likelihood value with the signal yield free in the fit, and $\mathcal{L}_{0}$ is that with the signal yield fixed to zero.
The signal efficiency determined from the signal MC sample is $\epsilon_{\rm sig} = (26.48 \pm 0.03)\%$.  
With these inputs, we obtain $\mathcal{B}(D^0 \to \gamma\omega) = (2.12 \pm 0.40) \times 10^{-5}$.
The reliability of the DNN-based method is further cross-checked by the traditional cut-based analysis~\cite{supplement}.

%------------------------------------------------------------------------------
The systematic uncertainties in the BF measurement are summarized in Table~\ref{tab:BF syst err} and discussed below.
With the DT method, uncertainties related to the ST selection cancel.
The uncertainty in the ST $\bar{D}^0$ yield reflects the uncertainty in the fit to the $M_{\rm BC}$ distributions and is assessed by varying the signal and background shapes.
The uncertainties in the tracking or particle identification (PID) of $\pi^\pm$ are studied with samples of DT $D\bar{D}$ hadronic events~\cite{BESIII:2018apz}. 
The uncertainties associated with the $\gamma$ and $\pi^0$ reconstruction are estimated by studying the control sample of $J/\psi\to\pi^+\pi^-\pi^0$ events~\cite{BESIII:2011ysp} and DT $D\bar{D}$ hadronic events~\cite{BESIII:2018nzb}, respectively.
The uncertainty due to the no extra track requirement is investigated with the control sample of $D^0\to K^-\pi^+\pi^0$ events. 
Following the previous studies~\cite{BESIII:2024mgg,BESIII:2024cbr,BESIII:2025biy}, we consider two primary sources of uncertainty in the DNN output, namely the ``model uncertainty'' and the ``domain shift''.
Model uncertainty is attributed to our imperfect knowledge about the network architecture.
The inherent stochasticity in the training process (e.g., random initialization, batch ordering, and dropout) can cause the model to converge to different local minima, resulting in variations in the predicted output.  
Following the model ensemble approach described above, this uncertainty is quantified by the relative error of the Gaussian-fitted BF distribution, which is determined to be 10.9\%.
On the other hand, domain shift describes the mismatch between datasets for training and inference, reflecting the potential data-MC inconsistency. 
Using a similar model training and fitting procedure, we measure the BFs of the control channels $D^0 \to \pi^+ \pi^- \pi^0$ and $D^0 \to \pi^0 \eta^\prime$, both of which are found to be in good agreement with the previous BESIII measurements~\cite{BESIII:2022qrs,BESIII:2018oqs}. 
The relative deviation of the measured BF for $D^0 \to \pi^+ \pi^- \pi^0$ from the published value is 0.4\%, which is assigned as the corresponding systematic uncertainty.
This consistency serves as a validation of our entire analysis framework, confirming its reliability across different final states.
The previous two uncertainties are combined in quadrature to give a total DNN output uncertainty of 10.9\%.  
The uncertainty of the fit method is assessed by modifying the baseline MC background shape and the fitting range.
First, to assess the shape uncertainty of the major background  $ D^0 \to \pi^+\pi^-\pi^0\pi^0 $ , we vary its yield by  $ \pm 30\% $ , corresponding to the quoted BF uncertainty of its dominant component  $ D^0 \to \omega\pi^0 $~\cite{PDG}.
Second, the $e^{+}e^{-} \to q\bar{q}$ background yields are changed by $\pm 5.4\%$ which corresponds to a change of $\pm 1\sigma$ in the measured cross section~\cite{Ablikim:2007zz}.
Finally, the low-side fitting boundary is varied by $\pm 0.05$. 
The uncertainty in the signal shape is already accounted for through the model ensemble approach.
The resulting variations in fitted signal yield of 2.5\%, 1.5\%, and 1.5\%, respectively, are combined in quadrature to obtain the overall fit method uncertainty of 3.3\%. 
The uncertainty associated with the quoted BFs of $\omega$ and $\pi^0$ decays is taken from Ref.~\cite{PDG}.
Assuming that all sources are independent, the total systematic uncertainty of the BF measurement is determined to be 12.1\%, by adding them in quadrature. 
With fit-related uncertainties considered, the final significance is conservatively estimated to be $6.0\sigma$.

\begin{table}[htbp]
\begin{center}
\caption{Relative systematic uncertainties on the BF measurement.}
\vspace{-1em}
\small
\begin{tabular}{l|c}
\hline \hline
Source & Uncertainty $(\%)$ \\
\hline
ST yield                             & 0.3\\
Tracking                             & 2.0\\
PID                                  & 2.0\\
$\gamma$ reconstruction              & 1.0\\
$\pi^0$ reconstruction               & 2.0\\
No extra track                       & 1.3\\
DNN output                           &10.9\\
Fit method                           & 3.3\\
$\omega$, $\pi^0$ BFs                & 0.8\\
\hline
Total                                &12.1\\
\hline
\hline
\end{tabular}
\label{tab:BF syst err}
\end{center}
\end{table}

\begin{figure}[htbp]\centering
\includegraphics[width=0.90\linewidth]{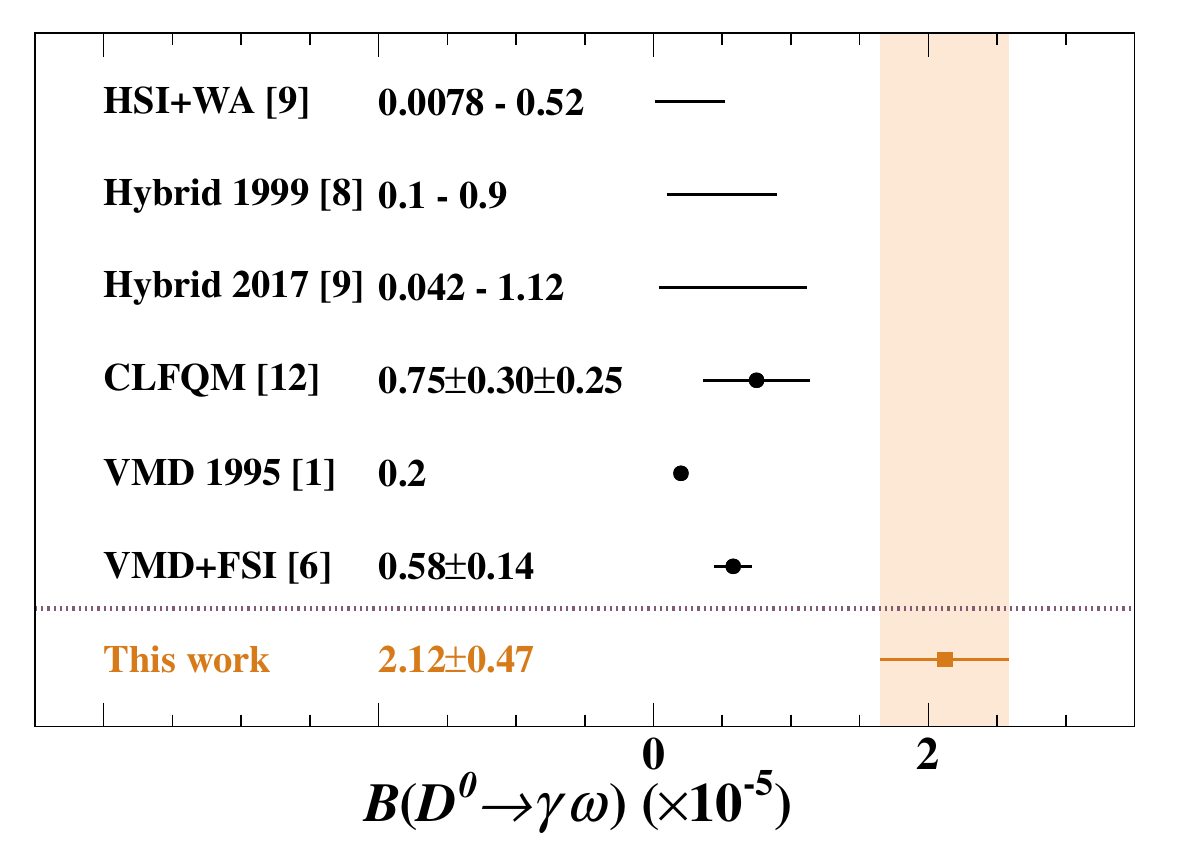}
\caption{Comparison of the measured $\mathcal{B}(D^0\to\gamma\omega)$ with long-distance theoretical predictions~\cite{deBoer:2017que,Fajfer:1998dv,Shen:2013oua,Burdman:1995te,Cao:2023gfv}. The short-distance contribution of ${\cal O}(10^{-9})$ would be indistinguishable from zero on this scale.  
Above the dotted line are the theoretical calculations, below is the experimental result.  
The orange band indicates the $1\sigma$ experimental uncertainty region. 
    }
    \label{fig:compare}
\end{figure}

%\section{Summary}
%------------------------------------------------------------------------------
In summary, using $e^+e^-$ collision data collected by the BESIII detector at a center-of-mass energy of 3.773\,GeV corresponding to an integrated luminosity of 20.3\,$\rm{fb}^{-1}$, we report the first observation of the radiative charm decay $D^0\to\gamma\omega$ with a significance of $6.0\sigma$. 
The absolute BF is determined to be $\mathcal{B}(D^0\to\gamma\omega)=(2.12\pm 0.40_{\rm stat}\pm 0.25_{\rm syst})\times 10^{-5}$.
This result is significantly larger than short-distance predictions, such as those from two-loop QCD calculations~\cite{deBoer:2017que}, unambiguously establishing the dominance of long-distance hadronic effects in radiative charm decays.  Also, as shown in Fig.~\ref{fig:compare}, this measurement lies systematically above other existing theoretical predictions. While CLFQM~\cite{Shen:2013oua} provides the closest prediction, it still exhibits tension, highlighting the need for further theoretical improvements.  
This result provides a crucial benchmark for discriminating among non-perturbative QCD approaches to $D \to \gamma V$ transitions.  
% {\color{red}Since this decay can be interpreted as proceeding via $D \to VV'$ followed by $V' \to \gamma$ conversion, the transverse polarization of the photon directly reflects the underlying helicity structure of the vector mesons. (delete?)} 

Using the PDG value $\mathcal{B}(D^0\to\gamma\rho^0)=(1.82\pm0.32)\times10^{-5}$ and the known masses of the $D^0,\omega$ and $\rho^0$ mesons~\cite{PDG}, 
we determine $\mathcal{D}^{\omega-\rho}$~\cite{Fajfer:2000zx} for the first time, $\mathcal{D}^{\omega-\rho} = 0.15\pm0.24$. 
% It is well below the 30\% difference threshold suggested in Ref.~\cite{Fajfer:2000zx}, showing no evidence of new physics.
Given the limited understanding of long-distance dynamics in these radiative charm decays, the extracted value helps to constrain theoretical predictions and motivates further theoretical investigations into the underlying dynamics.
% the BF ratio between $D^0\to\gamma\omega^0$ and $D^0\to\gamma\rho^0$ is calculated as $1.29\pm0.35$.
  
% Looking ahead, future high-statistics measurements of the photon polarization~\cite{Achasov:2023gey} would offer a sensitive probe of the decay dynamics and a potential signature of new physics~\cite{Biswas:2017eyn}.

%makes the radiative
%decays of charmed mesons $D\to \gamma V$, where $V=\omega, \rho$, an idea place
%for probing the contribution from the non-perturbative final-state interactions
%and searching for new physics.

%This 
%Again, it shows that the long-distance contributions
%from the VMD are dominant over the short-distance
%c → uγ dynamics. Meanwhile, the VMD contributions are
%still insufficient for accounting for the experimental data.

%%%%%%%%%%%%%%%%%%%%%%%%%%%%%%%%%%%%%%%%%%%%%%%%%%%%
% \include{acknowledgement_2026-03-11}
% \clearpage

The BESIII Collaboration thanks the staff of BEPCII (https://cstr.cn/31109.02.BEPC) and the IHEP computing center for their strong support. This work is supported in part by National Key R\&D Program of China under Contracts Nos. 2023YFA1606000, 2023YFA1606704, 2025YFA1613900; National Natural Science Foundation of China (NSFC) under Contracts Nos. 12635006, 11635010, 11935015, 11935016, 11935018, 12025502, 12035009, 12035013, 12061131003, 12192260, 12192261, 12192262, 12192263, 12192264, 12192265, 12221005, 12225509, 12235017, 12342502, 12361141819, 12535005; the Chinese Academy of Sciences (CAS) Large-Scale Scientific Facility Program; the Strategic Priority Research Program of Chinese Academy of Sciences under Contract No. XDA0480600; CAS under Contract No. YSBR-101; 100 Talents Program of CAS; 
the Excellent Youth Foundation of Henan Scientific Commitee under Contract No.~242300421044;
Joint Large-Scale Scientific Facility Fund of the NSFC and the Chinese Academy of Sciences under Contract No.~U2032104;
Zhongyuan Young Top Talent Program;
The Institute of Nuclear and Particle Physics (INPAC) and Shanghai Key Laboratory for Particle Physics and Cosmology; Agencia Nacional de Investigación y Desarrollo de Chile (ANID), Chile under Contract No. ANID CCTVal CIA250027; ERC under Contract No. 758462; German Research Foundation DFG under Contract No. FOR5327; Istituto Nazionale di Fisica Nucleare, Italy; Knut and Alice Wallenberg Foundation under Contracts Nos. 2021.0174, 2021.0299, 2023.0315; Ministry of Development of Turkey under Contract No. DPT2006K-120470; National Research Foundation of Korea under Contract No. NRF-2022R1A2C1092335; National Science and Technology fund of Mongolia; Polish National Science Centre under Contract No. 2024/53/B/ST2/00975; STFC (United Kingdom); Swedish Research Council under Contract No. 2019.04595; U. S. Department of Energy under Contract No. DE-FG02-05ER41374

\bibliography{References.bib}

\end{document}